\documentclass[printer]{aa}  

\usepackage{graphicx}
\usepackage{txfonts}
\begin{document}

%
   \title{TAPAS, a web-based service of atmospheric transmission computation for astronomy}


   \author{J.L.Bertaux
          \inst{1}
          \and
          R.Lallement
           \inst{2}
	\and
         S.Ferron \inst{3}
         \and
         C.Boone \inst{4}
          \and
         R.Bodichon\inst{4}
         }
   \institute{LATMOS,  University of Versailles Saint-Quentin,
              11 Boulevard d'Alembert, 78280 Guyancourt, France
              \email{jean-loup.bertaux@latmos.ipsl.fr}
         \and
           GEPI, Observatoire de Paris, 5 Place Jules Janssen, 92195 Meudon, France
             \email{rosine.lallement@obspm.fr}
	 \and ACRI-ST, BP234, 260 Route du Pin Montard,  06904 Sophia-Antipolis, France
	  \and IPSL, Institut Pierre-Simon Laplace, Place Jussieu, Paris, France\\
      }
   \date{Received October 2013}

 
  \abstract
{Spectra of astronomical targets acquired from ground-based instruments are affected by the atmospheric transmission.}
{The authors and their institutes are developing a web-based service, TAPAS (\textbf{T}ransmissions \textbf{A}tmosph\'{e}riques \textbf{P}ersonnalis\'{e}es pour l'\textbf{AS}tronomie,  or, \textbf{T}ransmissions of the \textbf{A}tmos\textbf{P}here for \textbf{AS}tromomical data). This service,  freely available, is developed and maintained within the atmospheric ETHER data center.}
{TAPAS computes the atmospheric transmission in the line-of-sight to the target indicated by the user. The user files a request indicating the time, ground location, and either the equatorial coordinates of the target or the Zenith Angle of the line-of sight (LOS). The actual atmospheric profile (temperature, pressure, humidity, ozone content) at that time and place is retrieved from the ETHER atmospheric data base (from a combination of ECMWF meteorological field and other informations), and the atmospheric transmission is computed from LBLRTM software and HITRAN data base for a number of gases: O$_{2}$, H$_{2}$O, O$_{3}$, CO$_{2}$, and Rayleigh extinction. The first purpose of TAPAS output is to allow identification of observed spectral features as being from atmospheric or astrophysical origin. The returned transmission may also serve for characterizing the spectrometer in wavelength scale and Instrument Line Spectral Function (ILSF) by comparing one observed spectrum of an atmospheric feature to the transmission. Finally, the TOA "Top Of Atmosphere" spectrum may be obtained either by division of the observed spectrum by the computed transmission or other techniques developed on purpose. The obtention of transmissions for individual species allows more potentialities and better adjustments to the data.}
{In this paper, we describe briefly the mechanism of computation of the atmospheric transmissions, and we show some results for O$_{2}$ and H$_{2}$O atmospheric absorption. The wavelength range is presently 500-2500 nm, but may be extended in the future.}
   { It is hoped that this service will help many astronomers in their research. The user may also contribute to the general knowledge of the atmospheric transmission, if he/she finds systematic discrepancies between synthetic transmissions and the observed spectra. This has already happened in the recent  past. The address is http://ether.ipsl.jussieu.fr/tapas/}

   \titlerunning   {TAPAS, on line atmospheric transmissions}

   \maketitle

\section{Introduction}
All astronomical targets (stars, galaxies, planets, exo-planetsâ etc) are seen from ground-based observatories through the Earth's atmosphere, which is polluting their spectra. In the field of stellar high resolution spectroscopy, there is a growing need for an accurate correction of atmospheric transmission, to reach interesting but contaminated spectral regions,
 to extract the best information from contaminated spectral regions, or simply make use of them instead of excluding them from analyses. A number of astronomers are adding to their target list a hot, bright star that is observed every night to be used as a template for the atmospheric lines, with a number of disadvantages: - loss of precious observing time; -the existence of many stellar features remaining in this template; - the impossibility of reaching the true continuum of the target because it is mixed with the one of the bright star; -the absence of feasibility in the case of multi-object spectrographs and weak targets. Today progresses in the molecular databases and radiative transfer models allow improved computations of the atmospheric transmission. Such a synthetic transmission has already been used in the past to correct for water vapor absorption (\cite{Lallement}), however a unique transmission spectrum computed for a standard atmosphere was used, adapting to the air mass and humidity by simply amplifying the lines of this unique spectrum. The transmission however is a function of the altitude and the atmospheric profile at the observatory. What we propose here is an online tool providing a state-of-the-art computation of this transmission, optimized for the location and time of the observation by the use of the actual meteorological field from ECMWF (European Center for Medium-range Weather Forecast).
 
There are several ways to use such synthetic atmospheric transmissions: -1) by visual comparison with the target spectrum, to identify telluric features; -2) by fitting the data to a model including the atmospheric transmission, with or without final adjustment to the average airmass and(or) water vapor evolution; - 3) by dividing the data with such an optimized transmission template. Importantly, our advice for an optimal use of the TAPAS tool, for the second and third purposes, is the downloading of the H$_{2}$O and O$_{2}$ absorptions spectra separately, which allows refined adjustments and taking into account differential behavior of the two species at the time of the observations.

\begin{figure*}
\centering
\includegraphics[width=12cm]{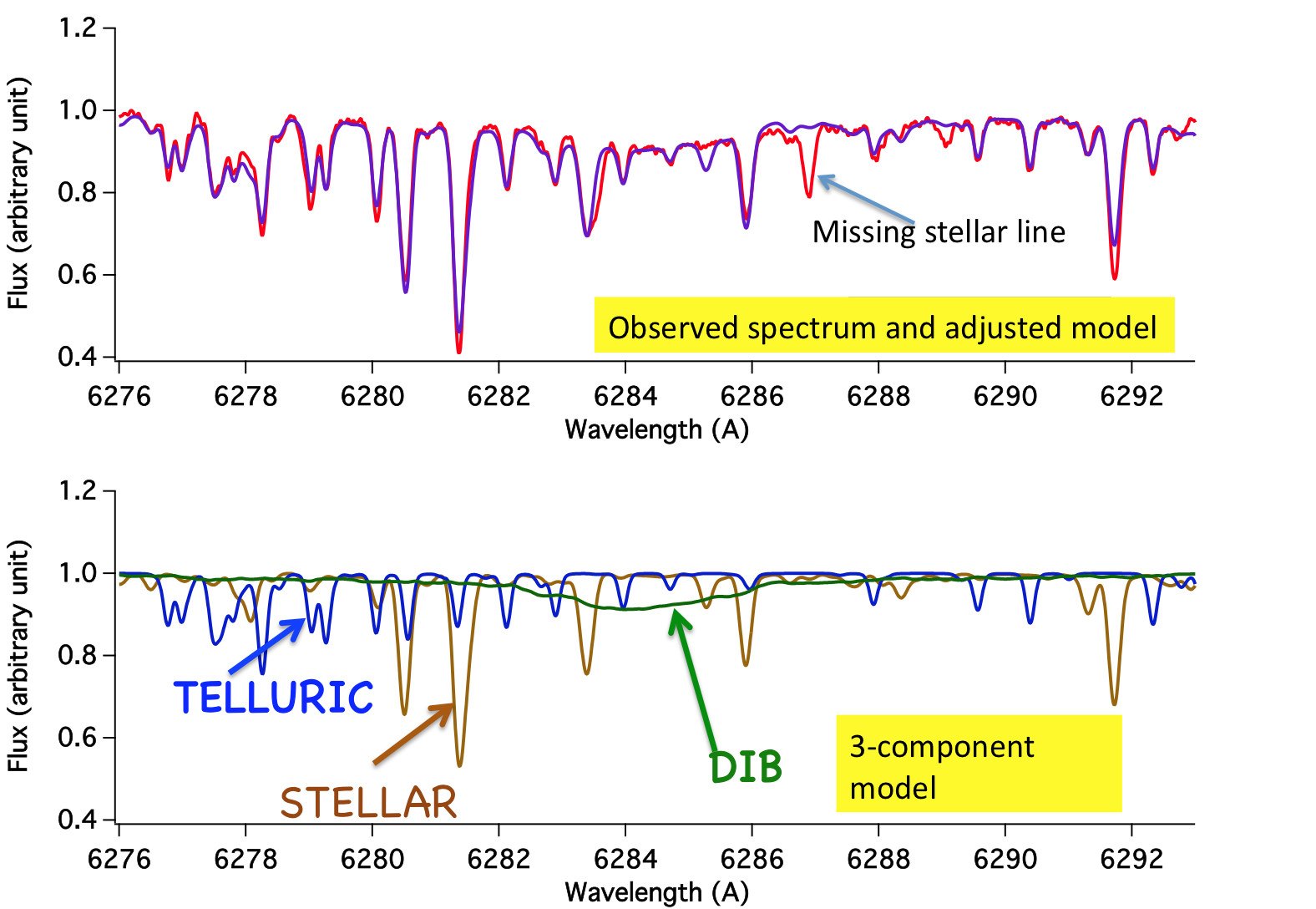}
\caption{Illustration of data-model adjustment using TAPAS spectra: a high-resolution (R=65,000) stellar spectrum is shown in the region of the 628.4 nm DIB (top). Data (in red) have been recorded from the Paranal Observatory with the FLAMES-UVES spectrograph at the VLT and adjusted to the convolved product (blue line) of three models shown at bottom: -a stellar model, computed for the appropriate stellar parameters (brown line), -a telluric transmission (blue) and a DIB empirical model (green). The use of the TAPAS transmission allows an optimal adjustment and estimate of the DIB equivalent width. It also allows revealing unambiguously missing lines in the stellar model, such as the feature at 628.7 nm (see (\cite{chen13} )).}
\label{dib_example}
\end{figure*}

An example of use of the second  method is shown in Fig.\ref{dib_example}  
The goal  was the extraction of a diffuse interstellar band (DIB), a feature that is imprinted by the intervening interstellar medium in astronomical spectra and is still the object of several studies aiming at identifying its carrier. In the case of the 628.4 nm DIB, strong molecular oxygen lines surround its absorption, which renders the measurement of its equivalent width (EW) difficult, especially for nearby objects (weak DIB), low resolution and large air mass. The O$_{2}$ transmission is computed with TAPAS for the altitude and location of the observatory (here the ESO Paranal and La Silla observatories in Chile). Fig. \ref{dib_example} shows a global adjustment to the data of a combination of stellar, telluric (TAPAS) and diffuse band models (\cite{chen13}). 
The choice between the different methods depends on the characteristics of the spectrum and the feature to be observed.
 
 In spectral regions highly populated with strong but narrow lines, one can also take advantage of the variable EarthÕ's velocity on its orbit. Looking at a star (not too far from ecliptic) at various periods, the comb of telluric lines is displaced, and most of the star spectrum may be reconstituted. This method, which was suggested and implemented by one of us (J.L. Bertaux) was first tested for the cool star beta Geminorum, allowing the identification of many molecular lines in the spectrum of the star in the range 930-950 nm  (\cite{Widemann}), where the telluric H$_{2}$O absorption is severe. The use of TAPAS would allow a better estimate of the depth of these star lines.
 
The interest of O$_{2}$ atmospheric lines as a wavelength standard for high precision measurements of star radial velocities (in search of exo-planets) has been also fostered and tested recently (\cite{Figueira}). They found that the stability of this atmospheric lines system as measured by HARPS spectrometer at La Silla was better than 10 m/s over 6 years (10 m/s is 3.3x10$^{-7}$ c, and corresponds to 1/100 of a pixel in HARPS), which is not surprising, because the wavelengths of the O$_{2}$ system are dictated by physical laws.

In the next section, we describe the principles of atmospheric transmission calculation used by TAPAS. Then are shown a few examples  of correction with TAPAS of  two stellar spectra of hot stars taken by two different spectrometers in two regions of the world, in  Sections 3 and 4 respectively for  O$_{2}$ and H$_{2}$O.  

\section{TAPAS computation: atmospheric profile and atmospheric transmission}

  \begin{figure*}
   \centering
  \includegraphics[width=\linewidth] {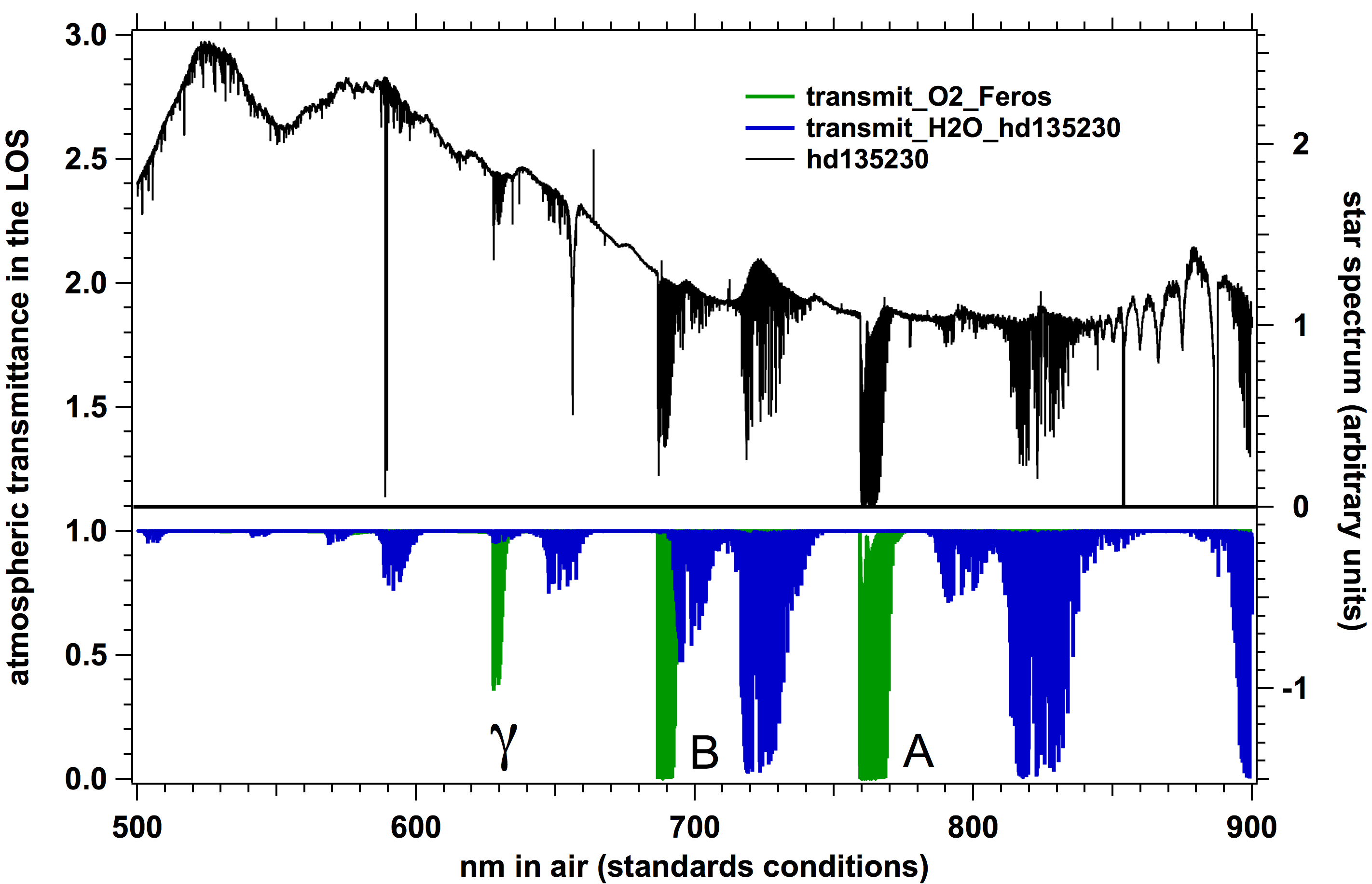}
  \caption{ Top panel: spectrum of the star HD135230 from La Silla, with spectrometer FEROS (ESO) associated to the 2.2 m MPIA telescope. Bottom panel: atmospheric transmittance calculated by TAPAS separately for H$_{2}$O (blue) and O$_{2}$ (green) in the range 500-900 nm, at the highest possible resolution with HITRAN. The atmospheric profile was computed for the time of the observation, integrated from the La Silla altitude and with the proper zenith angle. The location of O2 bands are indicated by letters, A, B, $\gamma$. The A band is also called the atmospheric band.}
 \label{FigGam}
\end{figure*}

In the visible or near Infra-red part of the spectrum, the lines of O$_{2}$, H$_{2}$O and CO$_{2}$ are the most conspicuous lines affecting the atmospheric transmission (see figure  \ref{FigGam} for O$_{2}$ and H$_{2}$O). Their local absorption depend mainly on the absolute concentration, but also on pressure and temperature. Oxygen is a well-mixed gas and the O$_{2}$ lines depend on the ground pressure, and on the vertical profiles of pressure and temperature. For H$_{2}$O, the integrated absorption depends also on the vertical distribution of H$_{2}$O concentration.
The TAPAS web-service is able to compute accurately the atmospheric transmission from a number of astronomical observatories.

TAPAS is one of the tools developed in the frame of ETHER, a French Atmospheric Chemistry Data Center maintained from CNES and INSU/CNRS fundings (www.pole-ether.fr). The database includes measurements acquired by satellite, balloons and aircraft or from other sources and the aim is to make them available to the entire scientific community for exploitation by modeling or assimilation. ETHER is also a vehicle for discussion and information on related scientific themes that may lead to greater comprehension and better use of existing data. 

A TAPAS user must be registered once, by filling a form on-line. This is done by logging to ether.ipsl.jussieu.fr/tapas/, then clicking on "request form". Once registered and getting a password, clicking on Òrequest formÓ will make appear the userÕs interface window 

If the simulated transmission is needed for the future (i.e, for observations planning purposes), TAPAS will use standard atmospheric profiles adequate for the season and latitude of the observatory, stored in the ETHER data base. The user may select a standard model among a list of 6 models, like ÒAverage latitude summerÓ.
If the simulated transmission is needed to correct a spectrum already acquired in the past, TAPAS will use the most realistic atmospheric profile (temperature T(z)  and pressure p(z) ) that is available. This atmospheric profile is an ETHER product called Arletty, which is computed by using the ECMWF (European Center for Medium-range Weather Forecast) meteorological field (analysis) for the date and time of the observation (actually, there is one field every 6 hours, and the one nearest in time to the observation time is selected). For atmospheric levels which are above the highest ECMWF field, Arletty is computing a profile by using the MSISE-90 algorithm (\cite{Hedin}). Arletty was developed for ETHER by ACRI company from the algorithms of Alain Hauchecorne at LATMOS.
The ECMWF data contain also the water vapor and ozone profiles. Water vapor is very important, because it contaminates the spectrum in many wavelength domains, with a large number of lines (figure \ref{FigGam} ).
Therefore, the input geolocation parameters that have to be entered by the TAPAS user are:

-the UTC time of the observation

-the location of the observer in longitude, latitude, and altitude (a list of selected observatories is available in a roll menu) 

-the RA and DEC of the astronomical target (RA, Right Ascension, DEC, Declination, in the J2000 system). From RA and DEC J2000, TAPAS computes the zenith angle, as an input parameter to the atmospheric transmission. Since an observer may wish to keep confidential its target (defined by RA and DEC), another available option is to give simply the zenith angle and the air mass will be computed by TAPAS. 

In order to compute the atmospheric transmission, TAPAS makes use of LBLRTM (Line-By-Line Radiative Transfer Model) ({\cite{Clough}) code and the 2012 HITRAN spectroscopic data base HITRAN (this is a high-resolution transmission molecular absorption database, www.cfa.harvard.edu/hitran/, ({\cite{Rothman}) for the HITRAN 2008 ).
This HITRAN database provides six of the key parameters for each line (or transition), namely the line position, the intensity, the air- and self-broadened half-widths, the temperature-dependence of the air-broadened half-width, and the pressure shift of the line. LBLRTM makes use of all theses parameters.

From the user-provided informations, TAPAS computes some other input parameters for LBLRTM , which are:

- the zenith angle

- the altitude of the observer

- the Arletty p-T vertical profile

- the H$_{2}$O and ozone profiles from ECMWF.

The atmospheric refraction, which bends the light path and increases slightly the atmospheric path is taken into account.

 Output parameters:
 
1. Selection of wavelength reference system. 
The user may select various systems:

- the wave number system  ( unit cm$^{-1}$, wave number= 10000/wavelength($\mu$m)). It has the advantage of being independent of the index of refraction of air.

- the wavelength in vacuum ( unit, nm, nanometer). It is also independent of the index of refraction of air.

- the wavelength (in nm) at 15$^{\circ}$ C and 760 mm Hg pressure (it is a commonly used laboratory reference wavelength standard; in particular, lines of Thorium Argon lamps which are frequently used to calibrate the wavelength scale of spectrometers are given in this particular wavelength reference system.)

It should be noted that the comparison of the observed spectrum, showing numerous O$_{2}$ or H$_{2}$O narrow lines, to the TAPAS spectrum, may be used to give an uttermost wavelength calibration of the observed spectrum. One thing that is not accounted for in the computation is the wavelength shift due to Doppler effect induced by the atmospheric wind. Actually, the ECMWF data contain this information and could be used in a future version of TAPAS. With a (high) typical velocity of 20 m/s, it would correspond to a displacement of 1/50 of a pixel for some of the highest dispersion spectrometers found on the astronomical market these days, like HARPS.

 \begin{figure*}
   \centering
   
   \includegraphics[width=\linewidth] {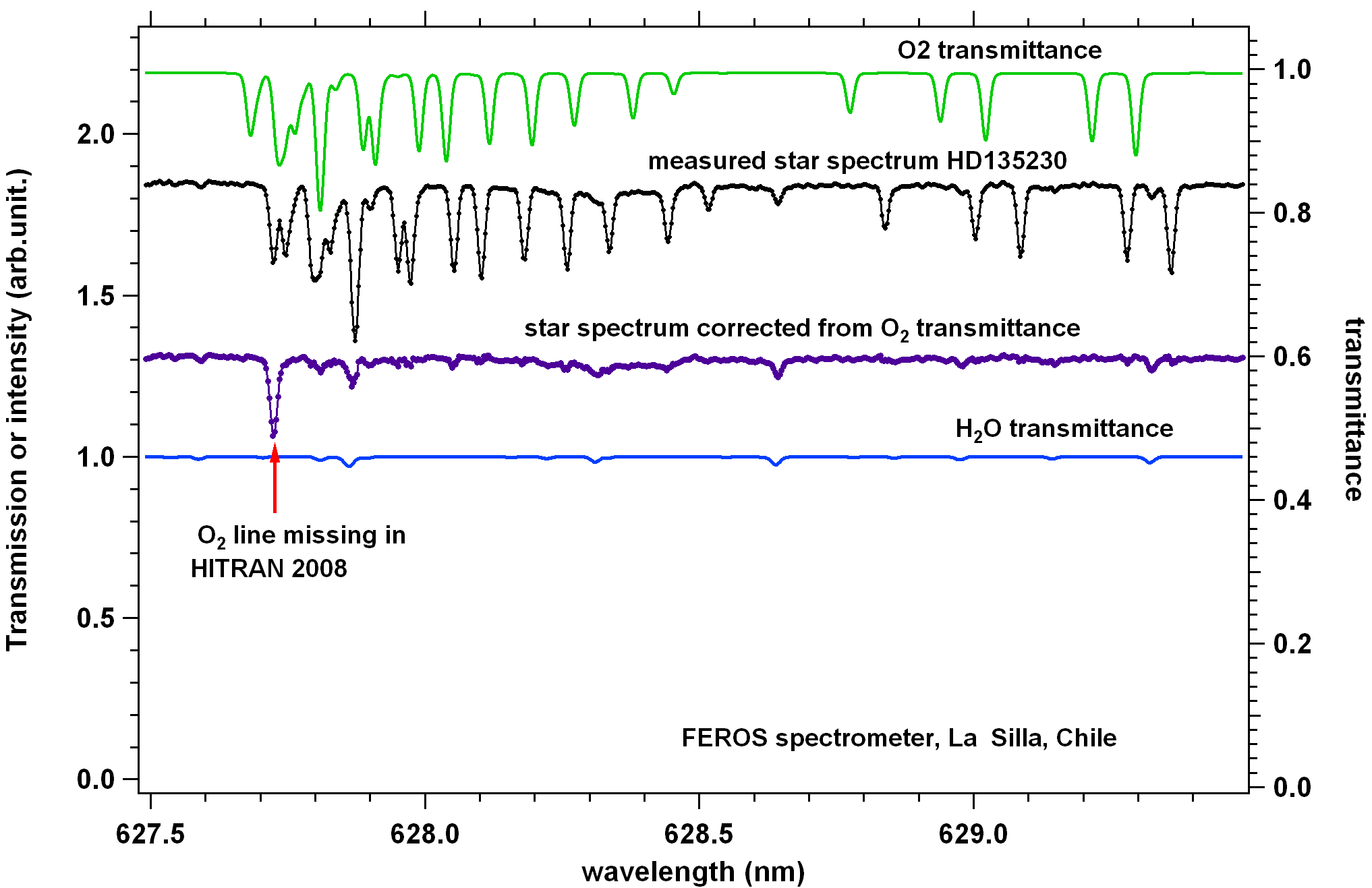}
  \caption{Green curve: the O$_{2}$ transmittance computed by TAPAS and convoluted for the appropriate resolution of 44,000, in a narrow spectral interval where O$_{2}$ lines are present (right scale). Black curve (left scale): the star spectrum showing absorption lines due to telluric O$_{2}$, shifted from the TAPAS transmission, because the FEROS wavelength calibration put the star spectrum in the Barycentric solar system reference. Dark purple curve: star spectrum divided by the convoluted transmittance, after proper wavelength shift (left scale, but vertically displaced for clarity). A conspicuous line remaining corresponds to an O$_{2}$ line absent from the HITRAN 2008 data base. Other small wiggles correspond to weak H$_{2}$O lines, as indicated by the light blue curve (H$_{2}$O only TAPAS transmittance, left scale).}
 \label{berv}
\end{figure*}

2. Definition of spectral interval for the computation. 
At present, the spectral interval may be selected within the window from 350 to 2500 nm.

3. Selection of atmospheric constituents:

The user may select the transmission computation within the following list:

- Rayleigh extinction, O$_{2}$, Ozone, H$_{2}$O. In the future, NO$_{2}$ and NO$_{3}$ from a climatology established from GOMOS ENVISAT (\cite{BertauxGomos}) measurements will be also available.
The user may also select the option in which the transmissions are calculated separately for each constituent, for a better identification of observed lines. The total transmission is then the multiplication of all individual transmissions. It also may be useful to get the H$_{2}$O transmission separately, because the quantity of H$_{2}$O may be adjusted by a power law of the transmission, T$^{X}$(H$_{2}$O), where X is the adjusting factor.

4. The Doppler effect and BERV option.

In the following we examine the spectrum of star HD135320 which was taken on February 11, 2009, at UTC 09:49:08, with the FEROS high resolution spectrometer  and 2.2 m MPIA telescope at La Silla, an ESO observatory in Chile, and displayed on Figure \ref{FigGam} (black). This star is a hot star, B9 spectral type, which contains only a limited number of relatively broad stellar spectral lines. The atmospheric transmittance for the relevant atmospheric profile was calculated by TAPAS separately for H$_{2}$O (blue) and O$_{2}$ (green) in the range 500-900 nm, at the highest possible resolution with HITRAN. The atmospheric absorption lines predicted by the TAPAS model are readily seen on the stellar spectrum.

We examine on figure  \ref{berv} a small 2 nm wavelength interval 627.5-629.6 nm where O$_{2}$ lines are prominent, by comparing first the star spectrum (black) with the TAPAS O$_{2}$ only transmittance (green), convoluted by a Gaussian profile with the FEROS resolution of 44,000. The O$_{2}$ atmospheric absorption lines are seen on the star spectrum, but there is an obvious wavelength shift between the two spectra. This is because the FEROS data reduction pipeline makes first an absolute wavelength calibration (with Thorium Argon spectral lines delivered by a dedicated lamp). Because of the rotation of the Earth around its spin axis, and because of the orbital motion of the Earth around the sun, there is a Doppler effect along the LOS affecting the observed star spectrum. The so-called BERV correction (Barycentric Earth Radial Velocity) is applied to the absolute wavelength of the spectrometer, to recover the exact spectrum which would be seen if the observatory were, not on Earth, but rather at the barycenter of the solar system. Since this barycenter has a galilean motion through the galaxy, all spectra of the same star may be compared to each other even if taken at various times and dates. Small variations of the star radial velocities are a major source of exo-planet discoveries, as predicted by (\cite{Connes}) and used later with great success (e.g.,\cite{Mayor}).

It must be emphasized that the Doppler effect is not a Doppler shift of the wavelength system, but rather a Doppler stretch. The wavelength $\lambda_{1}$ after barycentric correction is indeed computed from $\lambda_{0}$, absolute calibration of the spectrometer: 

\begin{equation}
														\frac {\lambda_{1}}{\lambda_{0}} = (1+\frac{BERV}{c} )   					
\end{equation}

in which c is the velocity of light, BERV=-Vr is the opposite of the radial velocity Vr= dR/dt of the Earth in the direction of the LOS. The convention is that BERV is positive when the Earth is approaching the star (more than the solar system barycenter), while Vr is negative. This change of sign between Vr and BERV is a source of confusion and has been unfortunately the source of many errors in the past. 
In the case of figure  \ref{berv} , BERV was positive (+ 30.249 km/s). Therefore, we had to ÒdebervÓ the star spectrum by applying an inverse correction (1-BERV/c) to get it in the observatory/spectrometer system, in which is also calculated the TAPAS transmission. Then, the ÒdebervedÓ star spectrum was divided by the O$_{2}$ only transmittance spectrum The result, which in principle represents the star spectrum outside the atmosphere (TOA), is displayed as a dark purple curve on figure  \ref{berv} . There is still a conspicuous line at 627.7 nm. In fact it is an O$_{2}$ line which was missing inadvertently in the HITRAN 2008 data base, but is present in the 2012 version (Rothman, private communication, 2012). Other small wiggles correspond to weak H$_{2}$O lines, as indicated by the dark blue curve (H$_{2}$O only TAPAS transmittance).
Instead of changing the star spectrum, one can also stretch the atmospheric transmittances with the same formula (1) that was used (presumably) by the spectrometerÕs data reduction pipe-line. TAPAS may do this for the user if desired, provided he/she fills the RA/DEC and time of the observation: included in TAPAS is the computation of BERV. However, this is only an option, because some pipe-lines do not apply the BERV correction before supplying the reduced spectrum to the observer.

\section{Atmospheric transmission of molecular oxygen O$_{2}$ }

\subsection{General}
In the following, we are using the TAPAS O$_{2}$ transmittance to correct by division a star spectrum. In addition to the Feros spectrum of HD135320, we used also a spectrum of another hot star HD 94756 taken with NARVAL spectrometer at Pic du Midi (France), with a greater resolution $\simeq$ 60,000. We have examined in details the three major bands of O$_{2}$ in the visible, namely A (also called Òatmospheric bandÓ), B, and $\gamma$  bands, displayed on figure \ref{FigGam}. The criterion of goodness is that, after division by the transmittance, the corrected star spectrum should be about flat. Of course, when a line is saturated (before instrumental convolution), there is no information on the star spectrum where the transmittance is 0, and there is no hope to retrieve it (except with the technique of several observations around the year, see above). However, in many cases, and even on not so saturated lines, we found remaining spectral features that may be assigned to an improper spectrometer wavelength calibration or an improper line width, or a non-gaussian ILSF profile. In some other cases, remaining features seem to be not connected to instrumental effects.

One caveat is in order here: our method of retrieval of the original spectrum, division by a convoluted transmission, is mathematically incorrect. The mathematically correct method would be to deconvolve the observed spectrum from the ILSF with a considerable wavelength oversampling, and then divide by the transmittance at the highest possible resolution. In practice, however, even with high SNR spectra, the data noise would not allow this mathematically correct method. This shortcoming of the division method is obviously more important with strong absorption features than with small ones.

We have used three versions of the HITRAN data base, hoping that the observations would show an improvement with the latest versions. In some cases, later versions are obviously better. In many other cases, we found that the remaining spectral features were much larger than the differences between the various O$_{2}$ HITRAN data bases. The present version of TAPAS is therefore based, for O$_{2}$, on the latest version contained in the HITRAN web site (December 2012) (except for the continuous absorption, as will be explained later).

The three versions are: 

short name HI08: this is the original HITRAN 2008 data base.

short name V10: HITRAN 2008 plus an update for H$_{2}$O, O$_{3}$, and for O$_{2}$ as discussed in Gordon et al. (2011).

short name V11: same as V10 plus a correction  for the O$_{2}$ A band of isotopologue $^{16}$O$^{16}$O. 

We found that the difference of TAPAS transmittance between V11 and V10 in the A band, after instrumental convolution, was 1.4 10$^{-3}$ at most along the A band. In contrast, in the gamma band we found differences of up to 0.02 between V10 and HI08.
It may be noted that Gordon et al.( 2009) used high resolution spectra of the sun, obtained with a Fourier transform spectrometer from TTCON network, at a very large air mass (zenith angle 82.45$^{\circ}$) and cold temperature (-23$^{\circ}$C) to minimize water vapour. They were able to adjust some parameters of the B band, in order to reproduce the observation. Their spectral resolution is 0.02 cm$^{-1}$, or $\simeq$ 700,000 for B band, much greater than ours. 
The NARVAL actual spectral resolution was estimated to be  $\simeq$ 60,000 by comparing the observed star spectrum around a strong and isolated O$_{2}$ line at 769.687 nm to the TAPAS transmittance computed for various resolutions. 

\subsection{Examination of star spectra divided by the theoretical transmission: the O$_{2}$ gamma band} 

The gamma band extends from 627.65 to $\simeq$ 633 nm, and is constituted of a series of relatively weak lines with a maximum depth of 23\% at the Feros resolution of 44,000. On figure \ref{Fig627Gam}  is examined the short wavelength part of the O$_{2}$ gamma band as seen by Feros spectrometer. The two versions HI08 and V10 of the transmittance are indicated, and the two corrected spectra with the two versions.

The corrected spectrum is much flatter for the V10 version, indicating a real improvement in the line intensity data base from HI08 to V10. There might still be some under correction. The remaining spectral feature in the V10 corrected spectrum at 627.8 nm is due to an H$_{2}$O line.
There might still be some under correction with the V10 model, green line. We can see that there is still some depressions at the places of some absorption features, which may be used to quantify the deficit of computed absorption. It is well known that the operation of convolution conserves the equivalent width (EW) of an absorption feature. In the regime of weak lines, it can be shown that the EW$_{c}$ of a line in the corrected spectrum is the difference EW$_{m}$-EW$_{s}$, where EW$_{m}$ and EW$_{s}$ are respectively the equivalent widths of the weak line respectively in the measured and the calculated spectrum. Therefore, the EW of the corrected spectrum should be zero if the absorption has been modelled correctly. A deviation from 0 suggest an improper line strength. The correct line strength is the one that would give EW$_{s}$=EW$_{m}$.


 \begin{figure*}
   \centering
\includegraphics[width=12cm]{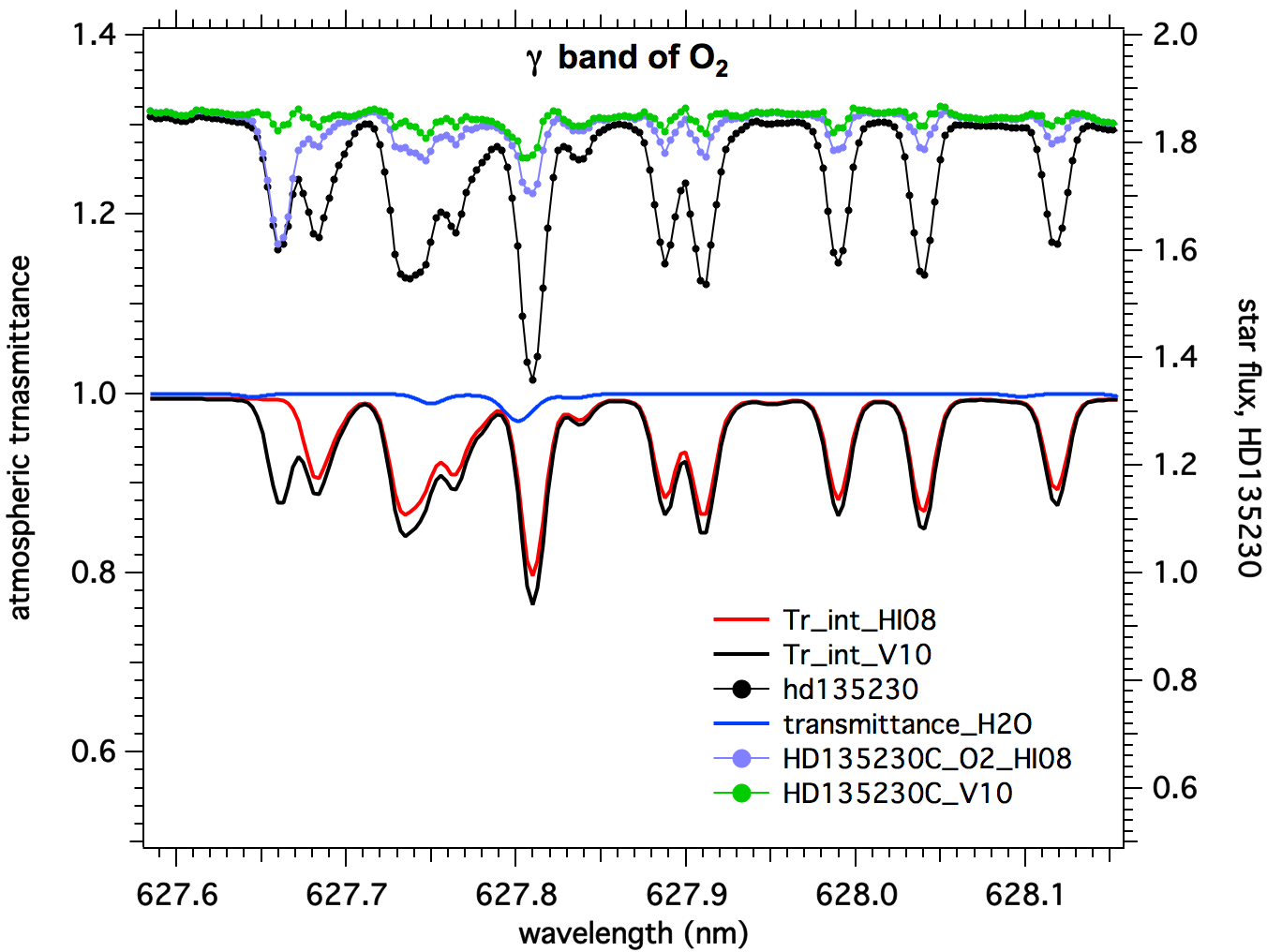}
\caption{Feros spectrum of star HD135230, before TAPAS correction (black dots and line, left scale in arbitrary units), and after correction by division of either the HI08 transmittance (red) or the V10 transmittance (black, left scale). There is a clear improvement of the V10 HITRAN O$_{2}$ data base over the earlier HI08 version, as indicated by the flatter green spectrum (corrected with V10) than the blue one (corrected with HI08). The H$_{2}$O transmittance is indicated by a blue line.}
 \label{Fig627Gam}
\end{figure*}

 \begin{figure*}
   \centering
\includegraphics[width=12cm]{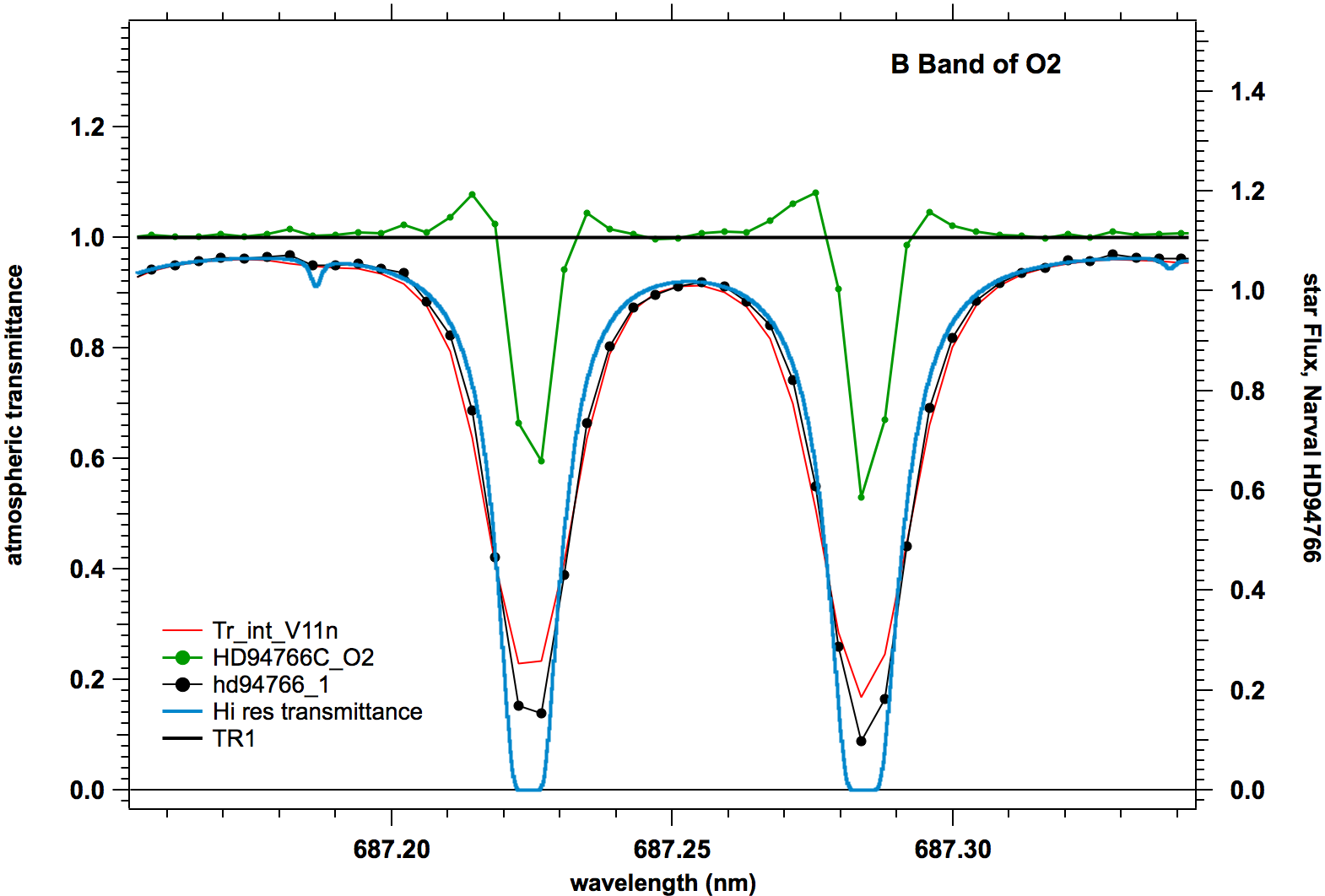}
\caption{A piece of Narval spectrum of HD94766  in the B band where two lines of O$_{2}$  showing shoulders and a self-reversal. The black line with dots is the original spectrum (right scale), the green line with dots is the corrected spectrum, after division by the convoluted transmittance. The red line is the TAPAS transmittance convoluted to a resolution of 66,000. The transmittance at high resolution is also plotted (blue), showing that both lines are saturated at center, and the exact value of the 4 central spectels of the corrected spectrum is somewhat meaningless.There are no H2O absorption features in this range. }
 \label{narvalspec2}
\end{figure*}

 \begin{figure*}
 \centering 
\includegraphics[width=12cm]{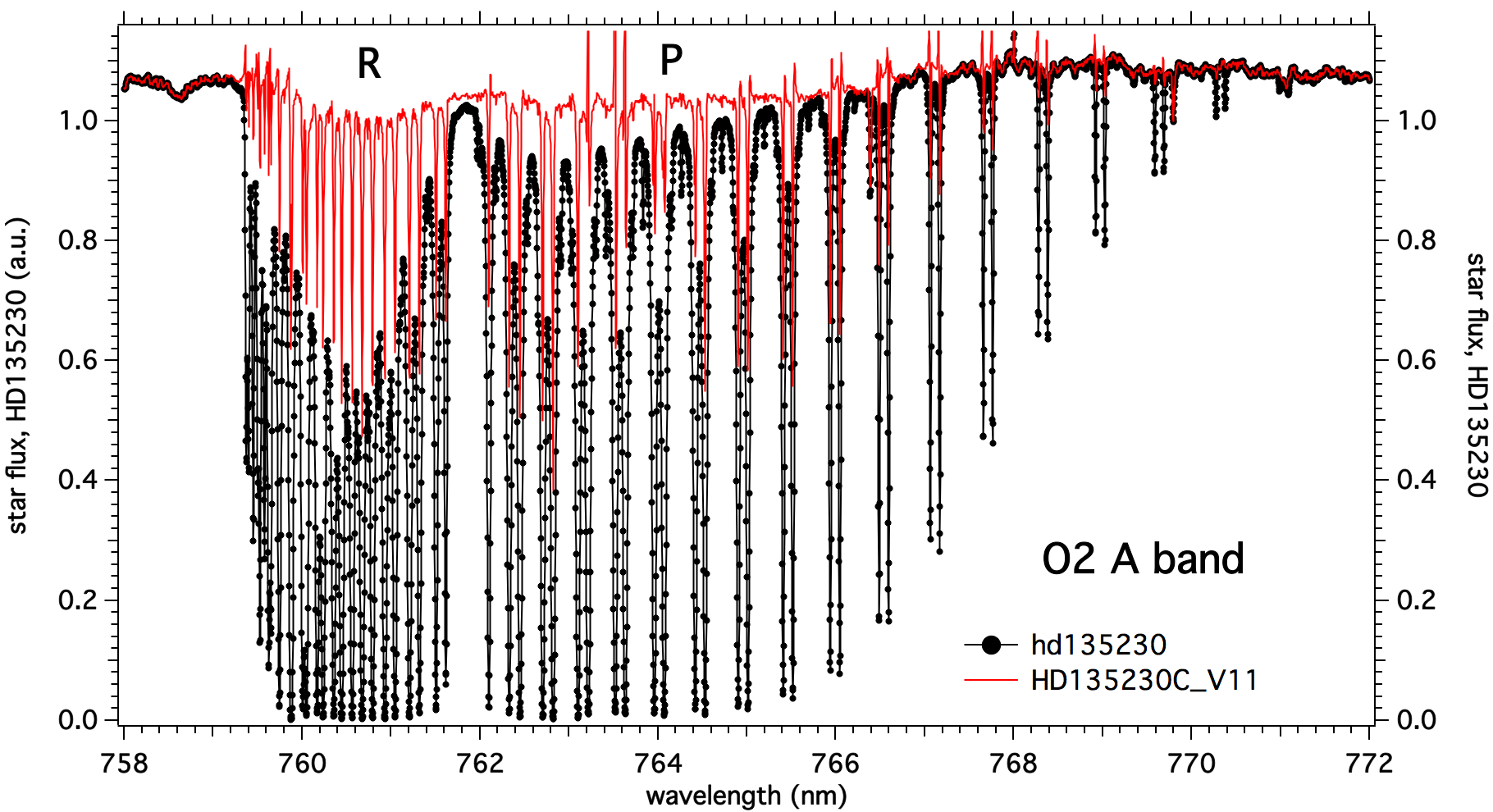}
\caption{The Feros spectrum of star HD135230 is displayed in the region of O$_{2}$ A band (black dots and line, a.u.), while the TAPAS-corrected spectrum is in red. The two branches P and R of the electronic transition are indicated. See text for discussion.}
 \label{Fig758O2}
\end{figure*}

 \begin{figure*}
   \centering
\includegraphics[width=12cm]{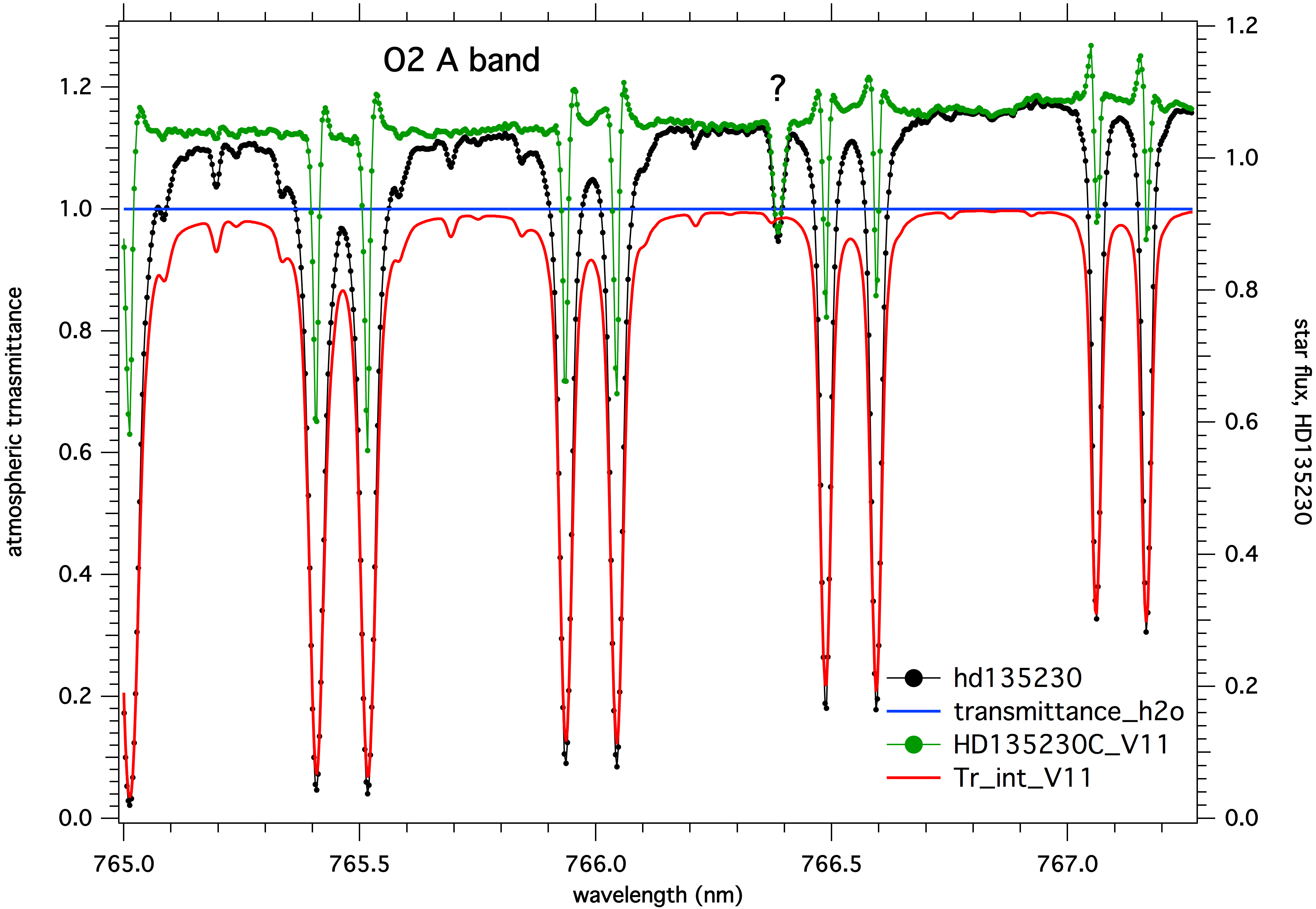}
\caption{A detail of the P branch in the Feros spectrum of HD135230: observed (black dots and points), and after correction with TAPAS (green points and line, right scale a.u.). There is a P Cygni pattern in all absorption features. The sign of the P Cygni changes along the spectrum. It goes to zero at 766.5 nm, then it reverses its sign, when going from one doublet to the next. There is a line at  766.39 nm in the star spectrum (question mark), not corrected by the atmospheric correction, identified as an interstellar line of Potassium at 766.4911nm. }
 \label{Figdetail}
\end{figure*}

\subsection{The O$_{2}$ B band} 

On figure \ref{narvalspec2}  is shown a NARVAL spectrum of two lines of the O$_{2}$ B band, before (black line) and after  (green line) correction by division of the transmittance (red line). Both lines are saturated at center, as indicated by the high resolution transmittance (blue line). 
On both lines, the corrected spectrum (green) has two shoulders and a strong self-reversal over four points. The fact that the short wavelength shoulder is higher than the long wavelength shoulder indicates a small mismatch between the wavelength scales of the transmittance from TAPAS/HITRAN and the observed spectrum. At present we put more confidence in the HITRAN scale than in the spectrometer scale. In order to revise the HITRAN scale, one would have to examine many spectra taken with various spectrometers, and only if a systematic shift is found, one could begin to question HITRAN. We believe that the HARPS spectrometer at La Silla would be the most reliable instrument for this exercise, because the search for exo-planets is the most demanding exercise in the field of wavelength calibration and stability.
At center of both lines, the self reversal over 4 ÒspectelsÓ certainly do not reflect the TOA star spectrum, but are the result of the saturated lines, where the TOA spectrum cannot be retrieved by definition. The self-reversal does not compensate for the shoulders to make EW=0, but the relationship of EW holding for weak lines does not apply to strong lines and nothing can be concluded on the true line strength from this non-0 EW. The correct way to check the line strength for strong lines would be to compare directly  EW$_{m}$ and EW$_{s}$ (computed with the instrument ILSF, and the same wavelength limits on the EW integral). 

While the shoulders (figure \ref{narvalspec2}) might be interpreted as an overestimate of this particular line strength in HITRAN, producing an over correction, we believe that they are rather a sign that the actual ILSF is slightly narrower in the wings than a gaussian shape with resolution 60,000, as was taken to compute the TAPAS spectrum. In any case, the corrected spectrum would reveal any stellar feature masked by O$_{2}$absorption, except for the four central spectels.

We have also noted that a Feros spectrum in the same region as Figure \ref{narvalspec2} show rather an overcorrection of lines (not shown here), except in a narrow interval, which coincides with a region where the star continuum is higher by 2-3 \%. Indeed, the original spectrum shows a small, but sharp increase at 687.41 nm that cannot be of stellar origin, and is instrumental: possibly an abrupt change in flat field correction, Dark Charge correction, Stray light removal, or change of grating order
This study case illustrates the potential use of TAPAS to characterize in great detail a spectrometer and its pipe-line, not only with the wavelength scale and the ILFS line shape, but also on small photometric features at the 2\% level or better.

\subsection{ The O$_{2}$ A band} 
Extending from ~759 to ~772 nm, this electronic transition of the homopolar O$_{2}$ molecule is the most conspicuous absorption band in the atmosphere of the Earth in the visible range. Named A band by Fraunhofer, it is also known as Òthe O$_{2}$ atmospheric bandÓ. Since O$_{2}$ is produced by photo-synthesis, it should be a prime target for characterizing exoplanets. Though this band is a nuisance for astronomers, it is heavily used in the frame of space borne Earth Observations, either to determine the cloud top altitude, or to determine the effective atmospheric path-length of solar photons scattered by the ground and sent back to space, sometimes after a few scatterings (Rayleigh and aerosols). This allows to ÒcalibrateÓ the path-length, and to determine the vertical column density of other molecules, like CO$_{2}$, as illustrated by the now flying TANSO/ GOSAT experiment, or the future OCO satellite. There are no H$_{2}$O lines in the spectral interval of the O$_{2}$ A band. Provided that the spectral resolution is good enough, the individual lines are separated except at some places in the R branch. Therefore, the use of TAPAS transmission spectra allows to identify the O$_{2}$ lines: all other features are genuine. When used in combination with the method of Bertaux (six-months interval star observations), most of the target spectrum may be 
retrieved, except for stars at high ecliptic latitude where the Doppler effect of the Earth's orbital velocity becomes negligible.

 We have tried the three versions of the data base; but the model differences are much smaller than some discrepancies between data and model. Therefore, we show here only the latest HITRAN version, V-11, which contained some corrections to the A band (\cite{Gordon}). 

The Feros star spectrum is displayed on Figure \ref{Fig758O2}  in the domain of the O$_{2}$ A band (black, Feros data). The overall-shape and structure, with the two separated branches P and R of the electronic transition, is readily captured on the original spectrum. The TAPAS-corrected spectrum is the red curve (with HITRAN version V-11). Many lines are saturated, and the retrieved values at saturated line-centers are spurious, and manifested by pseudo-absorption features in the corrected spectrum. Between the lines, the continuum is well retrieved, even in the R branch where lines are more densely packed. 

The upper envelope of the corrected spectrum is slightly depressed w.r.t. the star continuum outside the band, especially in the R branch. We believe that this is because TAPAS does not yet include an O$_{2}$ continuous absorption which was recently added to the HITRAN data base (Richard et al., 2012). This is due to Collision-induced absorption (CIA), and accounts for O$_{2}$-O$_{2}$ as well as O$_{2}$-N$_{2}$ and O$_{2}$-CO$_{2}$ collisions (\cite{Tran}).


On figure \ref{Figdetail} is displayed a detail of the P branch with 4 O2 doublets (Feros spectrum). We note first that all the weak lines other than doublets are well corrected (~flat spectrum after correction). The corrected spectrum doublet at 765.5 nm shows a so-called P-Cygni profile (a trough followed by a peak), meaning that there is a slight wavelength shift between data and model. However, there is a progressive change of sign of the P Cygni pattern over a small wavelength interval, from left to right. This is likely the sign of a problem in the wavelength calibration system of Feros, or its pipe-line, rather than a problem in the HITRAN data base. This may be due to a CCD engraving problem, which presents some irregularities (called ÒCCD stitchingÓ, Pasquini 2013, personal communication).

We note also that the positive peak is smaller than the negative peak in the P Cygni profile, indicating that the EW is $_{­}$ ­0. It could mean that these O$_{2}$ line strengths are actually larger than the HITRAN prediction. However, the absorption lines are strong, and we may be in a situation similar to that of Figure \ref{narvalspec2}, therefore one would have to consider the EW of data and TAPAS model to check the line strengths. 

There is also a line at 766.39 nm in the star spectrum, not corrected by the atmospheric correction. According to one of us (Rosine Lallement), it is an interstellar line of Potassium at 766.4911 nm. There is another one at 769.8974 nm that can be seen also in the same Feros spectrum (not shown here). The wavelength shift for the line of Figure  \ref{narvalspec2} is due to the motion of interstellar material and Doppler shift of 0.1 nm, corresponding to a heliocentric velocity of 39 km/s towards the sun. It illustrates how the use of TAPAS, by correcting the spectrum, and by comparing with the atmospheric transmission only, may help to discriminate between genuine astrophysical absorption features in the observation from atmospheric absorption features. It also illustrates the use of telluric lines for an accurate wavelength calibration, allowing an accurate determination of Doppler velocities of interstellar features, well below a small fraction of 1 km/s. The work of \cite{Figueira}  is in line with this statement.

\section{Water Vapor}
While O$_{2}$ is a well mixed gas in the atmosphere, the quantity of water vapor is highly variable in altitude, geography, season, and diurnal cycle. Being a tri-atomic molecule, its absorption spectrum is much more complicated than the O$_{2}$ spectrum (figure \ref{FigGam}).  In the range 0.5-1.0 $\mu$m, there are at many places relatively weak lines, and strongly saturated lines at other places. Because of its meteorological interest, weather forecast models and re-analysis are including water vapour as a measured, assimilated, and predicted parameter, as a function of altitude. Up to now, these models do not include the possibility of super saturation of water vapor, while tropospheric in-situ measurements do show sometimes such super saturation. 

Comparing TAPAS spectra with measured spectra do allow determining the quantity of H$_{2}$O in the atmosphere and comparing many observations to the ECMWF prediction could perhaps give some clues on super saturation. This is not in the scope of the present paper, where we compare the Feros spectrum to TAPAS output at various places between 500 and 920 nm, stopping short of the strong 936 nm band. 

As for O$_{2}$, the criterium for Ç goodness of fit Ç is that the star spectrum, after division by the calculated transmission, should be as flat as possible. On figure 8 is displayed the original star spectrum measured by Feros around the D1 and D2 sodium lines ~(589 nm),  compared to the H$_{2}$O transmission and divided by the transmission, convoluted with three possible spectral resolutions.
 It is seen that whatever is the resolution, the corrected spectrum is far from flat, showing a remaining absorption feature at all H$_{2}$O lines identified in the model. There is an overall deficit w.r.t. a straight line joining the continua on each side of the group of lines, which suggests that the actual quantity of H$_{2}$O is larger than predicted by TAPAS/ECMWF.

 \begin{figure*}
   \centering
   \includegraphics[width=12cm]{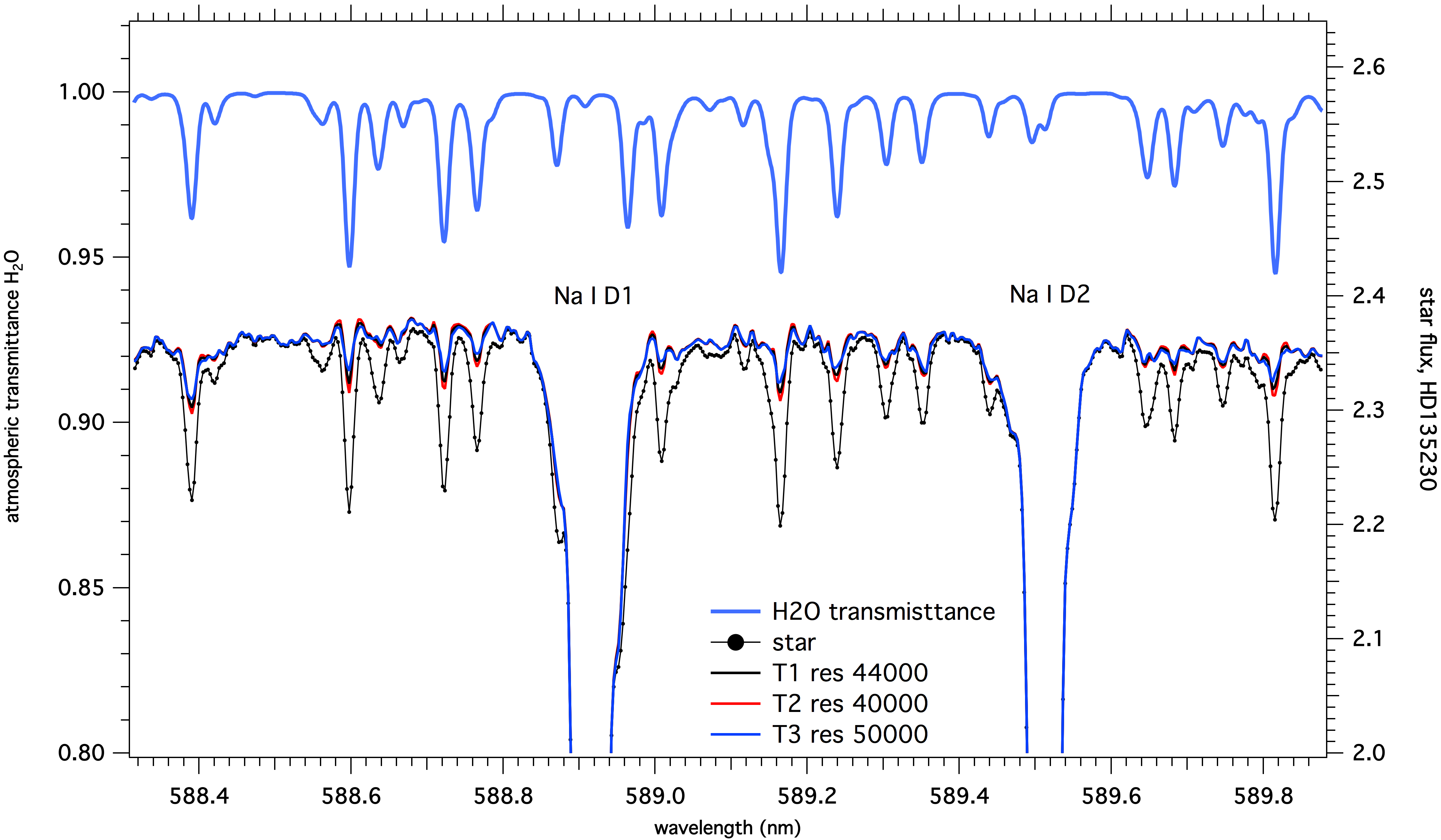}
\caption{Water vapor lines in the region of D1 and D2 sodium lines. The TAPAS transmission Spectrum is in blue, at top. All lines of the model are seen in the star Spectrum. The corrected spectrum has three versions, with 3 different resolutions. Whatever is the spectral resolution chosen to compute the TAPAS spectrum, there are residual absorption features that indicate that the quantity of H$_{2}$O in the model is lower than the reality. }
 \label{sodium}
\end{figure*}

\begin{figure*}
\centering 
 \includegraphics[width=12cm]{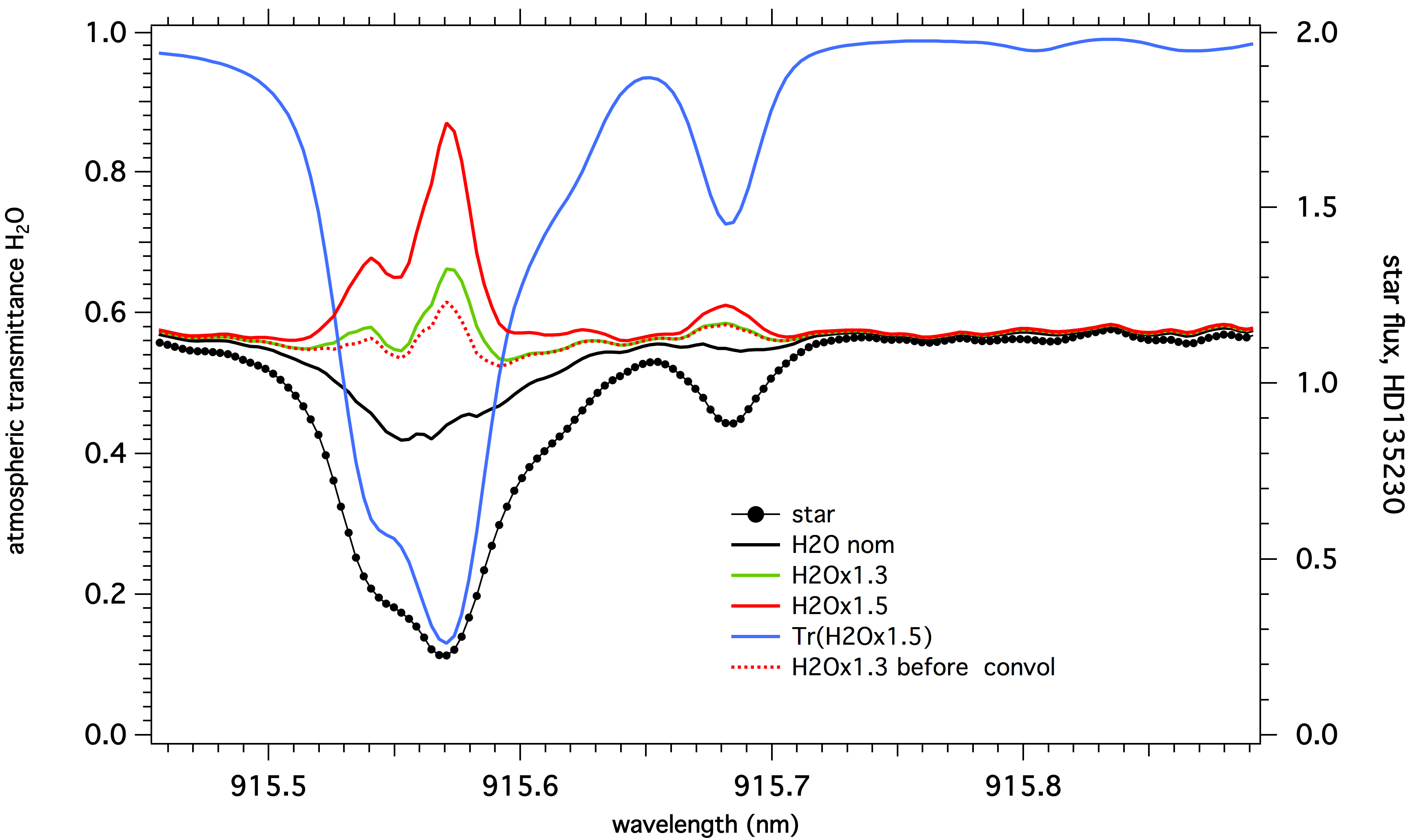}
\caption{Here, the wavelength shift is very small (no P cygni), therefore the wavelength scale is accurate. The solid black line is the TAPAS Ðcorrected with the nominal H$_{2}$O  column. While it is adequate on the H$_{2}$O  line at right, it is not enough for the lines on the left. See text for discussion. The red-dotted line is the exact calculation with 1.3 times the nominal H$_{2}$O column, while the green line is the approximate calculation. There is a significant difference at 915.57 nm where the observed H2O absorption is almost 85 \%.}
\label{FigGam2}
\end{figure*}

 \begin{figure*}
   \centering
  \includegraphics[width=12cm]{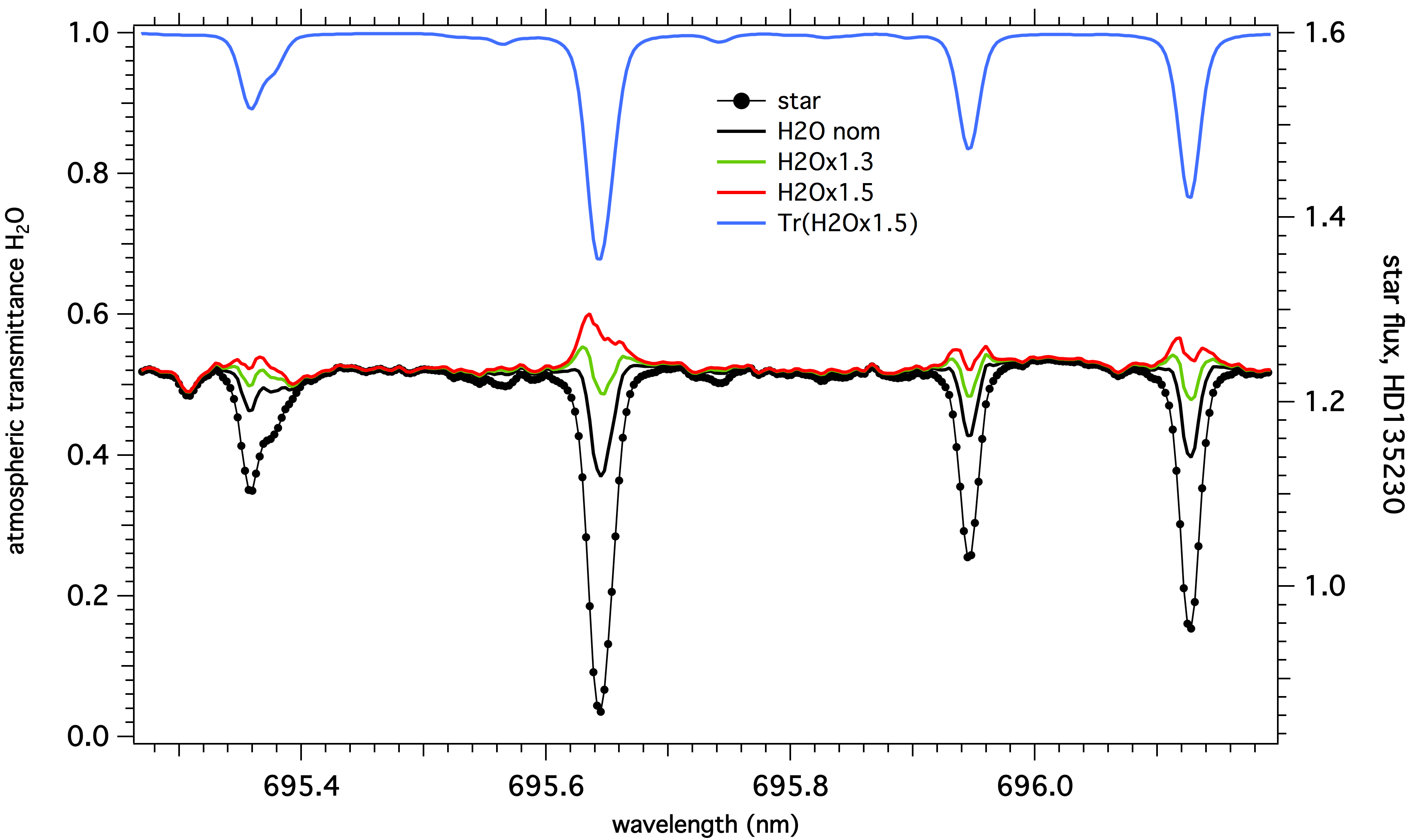}
\caption{The observed star spectrum (with dots where sampled) is corrected from the H$_{2}$O transmission computed by TAPAS with the nominal value of H$_{2}$O (black curve), and also with 1.3 (green) and 1.5 the nominal H$_{2}$O value. The remaining self-reversal is a sign of inadequate resolution/line width. Besides, even over this small spectral interval, it seems that the adequate factor to obtain a featureless spectrum after correction would need a different quantity of H$_{2}$O: while at 695.36 nm, the factor 1.3 would be approximately correct (green curve), it is too large at 695.64. }
 \label{FigvarH2O}
\end{figure*}



The presence of weak H$_{2}$O lines in this range has been a nuisance to discriminate interstellar weak sodium line from telluric H$_{2}$O absorptions. However these H$_{2}$O lines can be used to get an accurate wavelength scale, allowing the Doppler shift determination (\cite{Lallement}). In order to estimate what should be the transmission if the H$_{2}$O column was changed, we have used in the following a method that was used previously by \cite{Lallement} to correct even the weak H$_{2}$O lines. 
We have computed approximate estimates of the transmission by taking the TAPAS nominal transmission (thus, convoluted to the instrument resolution), elevated to the power 1.1, 1.3 or 1.5. This simulates the case in which the distribution of H$_{2}$O in the atmosphere would be the same as the one selected by TAPAS from ECMWF, multiplied by a factor 1.1, 1.3 or 1.5, constant at all altitudes.

This is only an approximation, because the correct calculation would be to use the TAPAS transmission spectrum at the highest resolution, to put it at the power X, and then to convolute to the instrument resolution. This correct calculation was done here for the X=1.3 H$_{2}$O case, and limited wavelength ranges, in order to check for the difference between exact and approximate calculations (figure \ref{FigGam2}). 
We have checked that for weak lines, the exact calculation gives similar results to the approximate calculation which consists of taking the power 1.3 of the convoluted spectrum computed with the nominal resolution. 
When the H$_{2}$O absorption is more severe, there is a more noticeable difference between the exact and approximate calculation of the corrected spectrum, as shown on figure \ref{FigGam2}. 
Therefore, the approximate calculation should be used with caution, and only with weak lines (say, giving less than 30\% absorption at a resolution of 40,000).


Keeping in mind that the convolution conserves the equivalent width EW,  therefore ignoring the self-reversals which are due to an inadequate resolution/ ILFS shape, the right value of X to fit the corrected star spectrum should display a EW=0 spectral feature for these relatively weak lines. 

Figure \ref{FigGam2} is an example where the same quantity of H$_{2}$O cannot fit two lines which are yet very nearby. It is clear that the observed absorption at 915.69 nm is rather well represented by the nominal H$_{2}$O value (black solid line), while this value is too small to represent well the absorption observed at 915.55 nm. 
Therefore, over very short wavelength intervals, the model absorption requires different amounts of H$_{2}$O to fit the observed absorptions, which is an unrealistic situation. 
A similar situation occurs in quite a different wavelength domain, around 695 nm as shown on figure \ref{FigvarH2O}. The TAPAS-corrected spectra of the same star with Feros are plotted with 3 values of the multiplying factor X : 1 (nominal case), 1.3 and 1.5. It is clear that that there is not enough water in the nominal case. 

A close examination of figure \ref{FigvarH2O} indicates that, even over this small spectral interval, the adequate factor X to obtain a featureless spectrum after correction would need a different quantity of H$_{2}$O:  while at 695.36 nm, the factor 1.3 would be approximately correct (green curve), it is too large at 695.64, and perhaps too small for the two other lines at 695.87 and 696.07 nm. This of course does not reflect reality: there is only one quantity of H$_{2}$O in the atmosphere. 




Therefore, over very short wavelength intervals, the model absorption requires different amounts of H$_{2}$O to fit the observed absorptions, which is an unrealistic situation. It may be due to shortcomings of the HITRAN data base. However, we lean toward a different explanation. First, we believe that at the time of Feros observation, the atmosphere was containing more water than predicted by ECMWF, by about 30\% $\pm$ 10\%. It may be a problem of rapid diurnal variation of the H$_{2}$O column. The ECMWF model is given every 6 hours, and Arletty/TAPAS is taking the nearest sample in time and geographical grid. Second, the actual vertical profile of H$_{2}$O concentration, temperature and pressure, may be different than ECMWF model. This may give rise to the observed discrepancies, because the lines parameters depend on these vertical profiles. Actually, it is conceivable that the series of observed absorptions of O$_{2}$ and H$_{2}$O on hot stars over the range 0.5-1.0 $\mu$m could be fitted by one single vertical profile of H$_{2}$O, p and T. This would open the future to a new series of atmospheric parameters which could be assimilated in the weather forecast models, if an adequate real-time pipeline is implemented for this purpose. This would require a better wavelength calibration and a better knowledge of the instrumental function, most likely variable over the observed wavelength domain. This would be a part of the pipe-line and fitting process. 

\section{Conclusions and perspectives} 
In this paper we have described the mechanism by which the TAPAS system provides a model of the atmospheric transmission to the user through a simple web interface. This was the main objective of the paper, and we hope that the users community will find useful this new service for their astronomical research. Their comments and suggestions will be greatly appreciated, and it is foreseen that TAPAS will evolve in response to the suggestions.
As a validation exercise of TAPAS, we have used two stellar spectra of hot stars taken by two different spectrometers in two regions of the world, focusing on the absorption of O$_{2}$ and H$_{2}$O, and corrected them with TAPAS transmission spectra. Our findings are as follows:

- all the lines predicted by TAPAS/ HITRAN are indeed present in the observed spectra (we have visually scanned the whole spectra with IGOR developed software). 

 - the lines observed in the star spectra are all coming from telluric absorption, except for some well known stellar lines, and some interstellar absorption lines.
 
 - there are some discrepancies in the wavelength scale of the lines (not described by a single shift) that we assign to the spectrometers pipe-lines.
 
- there are some discrepancies in the width and shape of individual lines, because the actual ILSF is different from the assumed Gaussian shape in TAPAS.

- for H$_{2}$O, the TAPAS/ ECMWF predicted column may be different : 30\% $\pm$ 10\% in our study case.

Therefore, TAPAS may be used for the following purposes by the general astronomer user working on a spectrum : 

1. identify the telluric origin (atmospheric) of one observed absorption feature, and assign more safely an astrophysical origin to the other lines not predicted in TAPAS/HITRAN.

2. identify the atmospheric absorbing gas molecule.

3. establish an accurate wavelength calibration scale of the observed spectrum in the reference system of the spectrometer.

4. determine the spectral resolution of the spectrometer (width of the ILSF).

5. determine the actual ILSF of the spectrometer. The actual ILFS of a spectrometer may be retrieved from a deconvolution of the observed spectrum in the region of a narrow line, as predicted from TAPAS for instance. A deconvolution may be regarded as the solution of a linear system greatly over-determined. We have tried both the method of least square fitting of the system, and the method of SVD (Singular Value Decomposition) as exposed and recommended by \cite{ruru}. Both methods are working well. 

6. correct the observed spectrum from telluric absorption. For H$_{2}$O, it may need some manipulation outside of the TAPAS environment,or several calls to TAPAS to get a good fit for all the lines.

7. a method for adjusting the H$_{2}$O column is proposed. 

TAPAS should be most useful on cool stars, where stellar lines are narrow, like O$_{2}$ and H$_{2}$O lines. Since these stars are heavily used in the search for exo-planets, it is particularly interesting to use the atmospheric lines for a wavelength standard (\cite{Figueira}). Also, if well corrected with TAPAS, some telluric contaminated regions which are presently discarded from the exo-planet search through radial velocity variations could be added to the analysis for a better retrieval of Vr changes.

In the future, TAPAS could be used for a better characterization of each astronomical hi res spectrometer, and an improvement of the associated pipe-lines (wavelength scale, Dark charge, sky light and stray light subtraction, spectrum extraction from a 2D image of a cross-dispersed spectrum). The actual ILFS could be determined in a number of spectral intervals, and provided with the spectrum as a separated file, or in a multi-file FITS format. In fact, several of these operations could be included in the reduction pipeline, with an automatic request to TAPAS. 

In our test-use of TAPAS, we have assigned most of the discrepancies to short-comings of the spectrometers, and not to inaccuracies of HITRAN (wavelength position and strength of lines, and sensitivity to temperature and pressure). It does not mean there are none; but in order to find them, it would be better to use HARPS spectrometer, which has a higher spectral resolution and an excellent wavelength calibration, due to the use of Thorium Argon lamps containing many lines. Also, the discrepancies should be found systematically for various spectrometers at various places and times, taking advantage of the constancy of line spectroscopy, universal and stable.

TAPAS will be improved and maintained by our team. Improvements will occur through corrections of mistakes, as signaled by the users. It is recommended that all users do report major mismatches of observed atmospheric lines with the TAPAS calculations, which may result into an improvement of the molecular spectra data bases. Also, we will add methane (CH4) and the O$_{2}$ continuum absorption following the line described in \cite{Richard}. Then absorption by NO$_{2}$ and NO$_{3}$ from a climatology established from GOMOS  on board the ESA ENVISAT spacecraft (\cite{BertauxGomos}) measurements may also be available.

As it has been signaled already, line parameters are influenced by temperature and pressure. Therefore, it is in theory possible to determine the vertical profile of H$_{2}$O , pressure and temperature, by inversing an observation (preferably a spectrum of a  hot star). This has been done already with success in the near IR ( \cite{Mumma}), in their search for methane on Mars. The method remains to be demonstrated for the 0.5-1 $\mu$m range, where a huge number of spectra are collected.  If the inversion is done in near real time, all these vertical profiles could be sent to weather forecast operational centers for assimilation in their models. Satellite borne instruments are doing this from space, but they suffer for the contribution of the surface, which is absent in the astronomical geometry, looking up. One thing that should be rather easy to retrieve would be the H$_{2}$O column, as we have shown in this paper. We suggest that this parameter is an extremely sensitive indicator of surface temperature, because of the Clausius-Clapeyron form of the law of water vapor saturation versus temperature. Therefore, monitoring the column of water vapor from various observatories could help to determine what is the real change of global surface temperature. Combined with the positive feed-back mechanism of water vapor (a green-house gas, which increase is increasing the green-house effect and therefore the temperature), it would address one major issue that the world has to face to-day.

\begin{acknowledgements}
This work is being supported by CNES (Centre National des Etudes Spatiales) and CNRS (Centre National de la Recherche Scientifique) and its Institut National des Sciences de lÕUnivers (INSU). It is a collaborative endeavor of LATMOS, GEPI, ACRI and ETHER/IPSL. We acknowledge the use of HITRAN data base and the LBLRTM radiative transfer code, the use of ECMWF data and the ETHER data center. We acknowledge useful discussions with Larry Rothman and Iouli Gordon, both in Reims and in Boston. Problems of TAPAS malfunctions should be addressed to Cathy Boonne (IPSL), cbipsl@ipsl.jussieu.fr.  Other questions should be addressed to one of following e-mails: jean-loup.bertaux@latmos.ipsl.fr, or rosine.lallement@obspm.fr. 
\end{acknowledgements}

\end{document}